\documentclass{aa}  

\usepackage{graphicx}
\usepackage{tablefootnote}
\usepackage{natbib}
\usepackage{txfonts}
\usepackage{xspace}

\newcommand{\um}     {~$\mu$m\xspace}

\newcommand{\cc}       {~cm$^{-3}$\xspace}

\begin{document}

   \title{Radio observations of evaporating objects in the Cygnus\,OB2 region}
%   \subtitle{I. Overviewing the $\kappa$-mechanism}

\author{N. L. Isequilla\inst{1}
 \and 
 M. Fern\'andez-L\'opez\inst{2}
  \and
   P. Benaglia\inst{1,2}
   \and 
    C.~H. Ishwara-Chandra\inst{3}
   \and
       S. del Palacio\inst{1,2}
   }
   \institute{Facultad de Ciencias Astronomicas y Geofisicas, UNLP, Paseo del Bosque s/n, 1900, La Plata, Argentina\\
              \email{nisequilla@fcaglp.unlp.edu.ar}
                \and Instituto Argentino de Radioastronomia, CONICET \& CICPBA, C.C.5, 1894 Villa Elisa, Argentina\
           \and National Center for Radio Astrophysics, TIFR, PB N3, Ganeshkhind, Pune 7, India
 }
 
   \date{Received Month XX, YYYY; accepted Month XX, YYYY}

% \abstract{}{}{}{}{} 
% 5 {} token are mandatory
 
  \abstract
   {We present observations of the Cygnus~OB2 region obtained with the Giant Metrewave Radio Telescope (GMRT) at the frequencies of 325~MHz and 610~MHz. In this contribution we focus on the study of proplyd-like objects (also known as free-floating Evaporating Gas Globules or frEGGs) that typically show an extended cometary morphology.
   We identify eight objects previously studied at other wavelengths and derive their physical properties by obtaining their optical depth at radio-wavelengths. Using their geometry and the photoionization rate needed to produce their radio-continuum emission, we find that these sources are possibly ionized by a contribution of the stars Cyg~OB2~\#9 and Cyg~OB2~\#22. Spectral index maps of the eight frEGGs were constructed, showing a flat spectrum in radio frequencies {in general}. We interpret these as produced by optically-thin ionized gas, although it is possible that a combination of thermal emission, not necessarily optically thin, produced by a diffuse gas component and the instrument response (which detects more diffuse emission at low frequencies) can artificially generate negative spectral indices. In particular, for the case of the Tadpole we suggest that the observed emission is not of non-thermal origin despite the presence of regions with negative spectral indices in our maps.}


   \maketitle
%
%-------------------------------------------------------------------

\section{Introduction}

The term protoplanetary disk (or, in short, proplyd) was introduced by \citet{1993ApJ...410..696O} to describe circumstellar disks around young low-mass stars, with the size of our own planetary system, revealed by Hubble Space Telescope images of the Orion Nebula (M42). 
These proplyds consist of a rotating neutral accreting disk with its surface ionized by an external source, usually a hot massive star. Hence, proplyds are usually found inside HII~regions.
 
Photoevaporation of a circumstellar disk due to nearby young massive stars is an efficient mechanism for molecule and atom destruction, resulting in an important mass loss and producing ultra compact HII~regions {\citep{1994ApJ...428..654H}}. Photoevaporation is produced either by EUV photons ($h\nu > 13.6$~eV) that ionize the gas and rise its temperature, or by UVL photons (6~eV$ < h\nu < 13.6$~eV) that dissociate molecules and also heat the gas. In 1998, Johnstone et al. presented a model to explain the photoevaporation of circumstellar disks by an external source of ultraviolet radiation and applied it to the Orion proplyds. Their model regards the sound speed in the ionized gas and the escape velocity of the gas from the protostellar gravitational field to find out the shape of the heated ionized material from the disk surface, which finally forms a thermal disk wind with a cometary structure.

Proplyds have been observed in Orion at different wavelengths \citep{1993ApJ...410..696O,1998AJ....116..322H,2002ApJ...570..222G,2005AJ....129..355B,2008AJ....136.2136R,2012ApJ...757...78M,2014ApJ...784...82M,2016ApJ...826L..15K,2016ApJ...826...16E}, as well as in other molecular clouds, such as the Carina Nebula \citep{2003ApJ...587L.105S}, NGC\,3603 \citep{1999AAS...194.6808B,2002ApJ...571..366M}, the Lagoon Nebula \citep{1998AJ....115..767S,2014ApJ...797...60M}, NGC 2024 \citep{2002hst..prop.9424S}, Sgr\,A* \citep{2015ApJ...801L..26Y}, Cygnus\,OB2 \citep{2012ApJ...746L..21W,2012ApJ...761L..21S,2014ApJ...793...56G,2016A&A...591A..40S} for instance.
 Their characteristic sizes range from 40~AU to 1000~AU \citep {1998AJ....116..854B,1999AJ....118.2350H,2012ApJ...761L..21S}. The masses of the ionized disks are similar to those measured in the non-ionized protostellar disks in Taurus and Ophiuchus molecular clouds (typical masses ranging between 0.003 -- 0.07~M$_{\odot}$, \citealt{2007ApJ...659..705A,2010ApJ...725..430M}), with the only exception of the disks within 0.3~pc to $\theta^1$Ori\,C that show a slightly lighter mass distribution.
 
 In the Cygnus\,OB2 region, \citet{2012ApJ...761L..21S} defined a new kind of objects called frEGGs (free-floating Evaporating Gas Globules). Like proplyds, these objects are also externally ionized and present extended cometary morphologies. The main difference is that proplyds harbor protostars in more evolved stages with protoplanetary disks already formed, whereas frEGGs are found to have larger molecular masses, indicating an earlier evolutionary stage in which the protostar is still undergoing strong accretion. The frEGGs are like the Evaporating Gas Globules (EGGs) defined by \citet{1996AJ....111.2349H}, but found in isolation, detached from their parental molecular clouds. In addition, frEGGs are about 10 times larger than proplyds (frEGGs are typically greater than $20\,000$~AU), so they can contain not just one but several young protostars with their circumstellar disks \citep{2012ApJ...751...69S}. FrEGGs are also more massive than Orion proplyds with total molecular gas masses exceeding \mbox{1--2~M$_\odot$} and contain molecular material inside the photoionized region. They are usually at further distances from their ionizing source than those in Orion.
 Previous radio observations on frEGGs toward NGC\,3603 and G5.97--1.17 \citep{2002ApJ...571..366M,2014ApJ...797...60M} found that some of these objects present negative spectral indices, associated with non-thermal emission. The non-thermal radiation from these ionized sources implies the presence of a population of relativistic electrons. Such particles can either be locally accelerated in the source or come from nearby cosmic-ray sources. 
 In the case of G5.97--1.17, \citet{2014ApJ...797...60M} suggest that electrons are accelerated at the shock produced by the  {photoevaporating} flow. 
 In this contribution, we present 325 and 610~MHz GMRT continuum observations toward the Cygnus\,OB2 region and report on the characteristics of the population of externally ionized proplyd-like objects also known as frEGGs. The paper is structured as follows. We review previous studies on these objects located in Cygnus in Section 2. In Section 3 we describe how the interferometric observations were taken and calibrated and the process for image production. We present the results in Section 4. In Section 5 we interpret the results and finally we summarize our main findings in Section 6.

%--------------------------------------------------------------------
\section{Proplyds and frEGGs in Cygnus}

 Cygnus is a large, northern-sky region of active star formation at an average distance of 2.5~kpc \citep{2008hsf1.book...36R}, which hosts nine stellar associations and a dozen of bright clusters. Nearly at the center of this complex region lies the Cygnus\,OB2 association,  {probably one of the most massive associations in the Galaxy}. The distance to the OB2 association is 1.4~kpc \citep{2012A&A...539A..79R}. Cygnus\,OB2 contains pulsars \citep{2003MNRAS.345..847B,2009ApJ...705....1C}, supernova remnants \citep{2013PASJ...65...14K,2017A&A...606A..14B} and binary stellar systems \citep{2004A&A...424L..39D,2015MNRAS.451..581L}, among other non-thermal radio-emitting sources. Recently, using IR and H$\alpha$ observations, a sample of ten externally-ionized objects has also been identified toward this region (\citealt{2012ApJ...746L..21W}, hereafter W12). Most of them are tadpole-shaped with a bright ionization front head roughly pointing to the central cluster of massive stars, and a tail extending in the opposite direction.

 A more recent study with Herschel FIR observations at 70~$\mu$m in the Cygnus region \citep{2016A&A...591A..40S} classified several objects ionized in their peripheries by external UV radiation as pillars, globules, evaporation gas globules and proplyd-like objects, measuring densities and temperatures for all of them. The EGGs and proplyd-like objects are 0.1--0.2~pc in size, with masses of 10-30~M$_{\odot}$, densities of 2.2--15.0$\times10^4 {\rm cm}^{-3}$ and average dust temperatures of 17~K. They also show a clear orientation toward the center of the association Cygnus\,OB2. 

 Perhaps the most studied of these proplyd-like objects is IRAS\,20324$+$4057, dubbed the Tadpole \citep{2012ApJ...761L..21S}, toward which there is a battery of observations at different wavelengths (W12; \citealt{2012ApJ...761L..21S,2014ApJ...793...56G,2016A&A...591A..40S}).
 IRAS\,20324$+$4057 is located at $RA$ = 20:34:13.27, $Dec$ = 41:08:13.8 [J2000], about $15'$ (6~pc at a distance of 1.4~kpc) from the center of the Cygnus\,OB2 association. It consists of an extended cometary nebula (7.7$\times10^4$~AU) oriented East-West, brighter and well defined in the northern edge and with no apparent axial symmetry, as seen in optical wavelengths \citep{2012ApJ...761L..21S}. Its morphology and orientation suggest a structure formed by the dynamic pressure of wind passing from a distant source or sources located west of the object, presumably a stellar wind from Cyg\,OB2\,\#8.
 \citet{2012ApJ...751...69S} observed the Tadpole using the VLA at 22~GHz and 8.5~GHz. They derived a negative spectral index suggesting the presence of non-thermal emission. 
The confirmation of non-thermal radiation associated with frEGGs would suggest that these objects are possible  accelerators of cosmic-rays, and perhaps even putative gamma-ray sources. Hence, frEGGS could become new laboratories for the investigation of particle acceleration processes in slow ($< 10^3$~km s$^{-1}$) flows, and possibly also as counterparts of unidentified gamma-ray sources.
 Because of their non-symmetries, large sizes and distances between the head and tail, these objects are different from those detected in the Orion Nebula.

%------------------------------------------

\section {GMRT observations}

The field of IRAS\,20324$+$4057 was observed with the Giant Metrewave Radio Telescope (GMRT)\footnote{www.gmrt.ncra.tifr.res.in} in Nov 2013 at the frequency bands of 325~MHz and 610~MHz, during 6~h and 3~h, respectively. The phase center was set at $RA$~=~20:32:50, $Dec$~=~41:16:50 (J2000), near the center of the Cyg\,OB2 association. The source 3C48 was used as the flux calibrator, and 2038$+$513 and 2052$+$365 were monitored as phase calibrators at 325 and 610~MHz respectively.

The data reduction and imaging were performed with the Astronomical Imaging Processing System (AIPS, \citealt{2003ASSL..285..109G}) following standard procedures; see for instance \citet{2015MNRAS.451...59M}. Also as in \citet{2015MNRAS.451...59M}, in addition, we applied a factor for the system temperatures at each frequency bands, to correct for the emission of the Galaxy.  The factors, $2.2\pm0.22$ for the 325~MHz band and $1.3\pm0.09$ for the 610~MHz band, were derived from data taken in Jan 2018, exclusively to this purpose (see \citealt {2015MNRAS.451...59M}, Appendix A, for the characteristics of the corrective process). 

We obtained a 325~MHz image of the field with a synthesized beam of $7.81'' \times 6.60''$, an rms of 0.2 mJy beam$^{-1}$, and an intensity peak of 0.270 Jy beam$^{-1}$. The corresponding image at 610~MHz was built with a synthesized beam of $7.60'' \times 5.97''$, resulting in an rms of 0.2 mJy beam$^{-1}$, and an intensity peak of 0.201 Jy beam$^{-1}$. 

The imaging step produced a phase shift -specifically during the self-calibration stage- between the intensity peak at both frequencies. This was corrected by selecting several random point sources (radio galaxies preferentially) across all the field of view, determining the coordinates of the intensity peak at both frequencies, computing the average offset between them, and applying this positional shift to the 610~MHz image  { ($RA_{\rm offset}=3.19 \pm 1.39 \arcsec, Dec_{\rm offset}=0.34 \pm 0.99 \arcsec$) }. This step was crucial before producing spectral index images, because combining the offseted images produced a gradual spectral index gradient with the same orientation as the positional shift in all the sources detected in the field of view. 

 {To generate the maps of spectral indices in MIRIAD, we convolved and regridded the 610 MHz image to exactly the same grid as in the 325 MHz one (thus the pixel size in both maps is the same, $1.5'' \times 1.5''$).}

%-------------------------------------- Two column figure (place early!)
  \begin{figure*}
   \centering
   \includegraphics[width=\hsize]{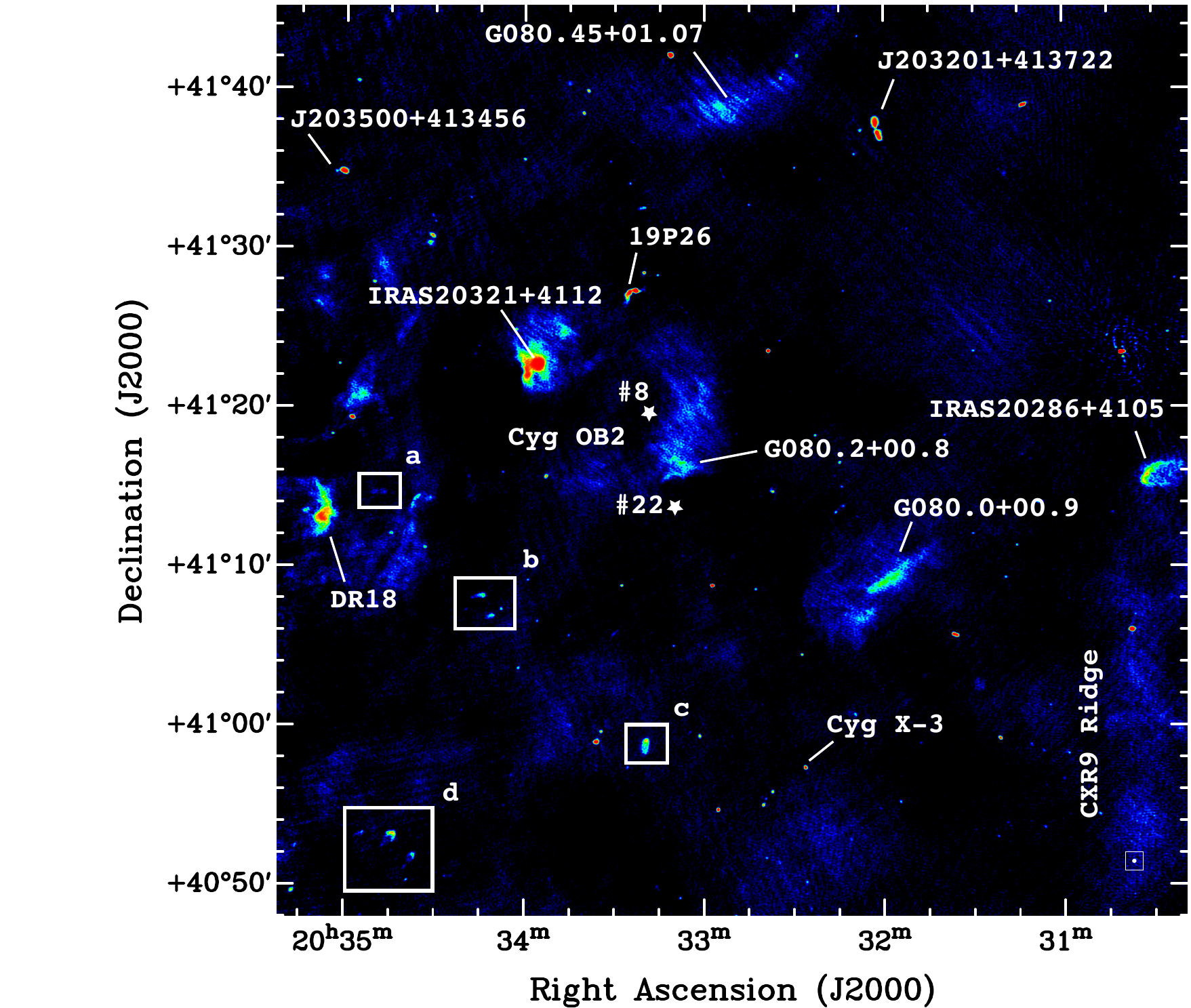}
   \caption{Cygnus OB2 image at 325 MHz; the synthesized beam is $7.81" \times 6.60"$, $P.A.$ = 88.6 degrees, the rms is 0.2 mJy beam$^{-1}$, and the intensity peak is 0.270 Jy beam $^{-1}$. The boxes in the figures show the regions where the proplyd sources studied here lie: source WDDGGHK 10 in box a, WDDGGHK 7, 8, 9 in box b, WDDGGHK 6 in box c, WDDGGHK 2, 3, 4 in box d.  {The white stars mark the position of putative ionizing agents}.}
   \label{Ffov}%
  \end{figure*}
%-----------> The font size is a bit out of proportion. Is it too much trouble to resize it or reduce the image size a bit?

 \begin{figure*}
   \centering
   \includegraphics[width=\hsize]{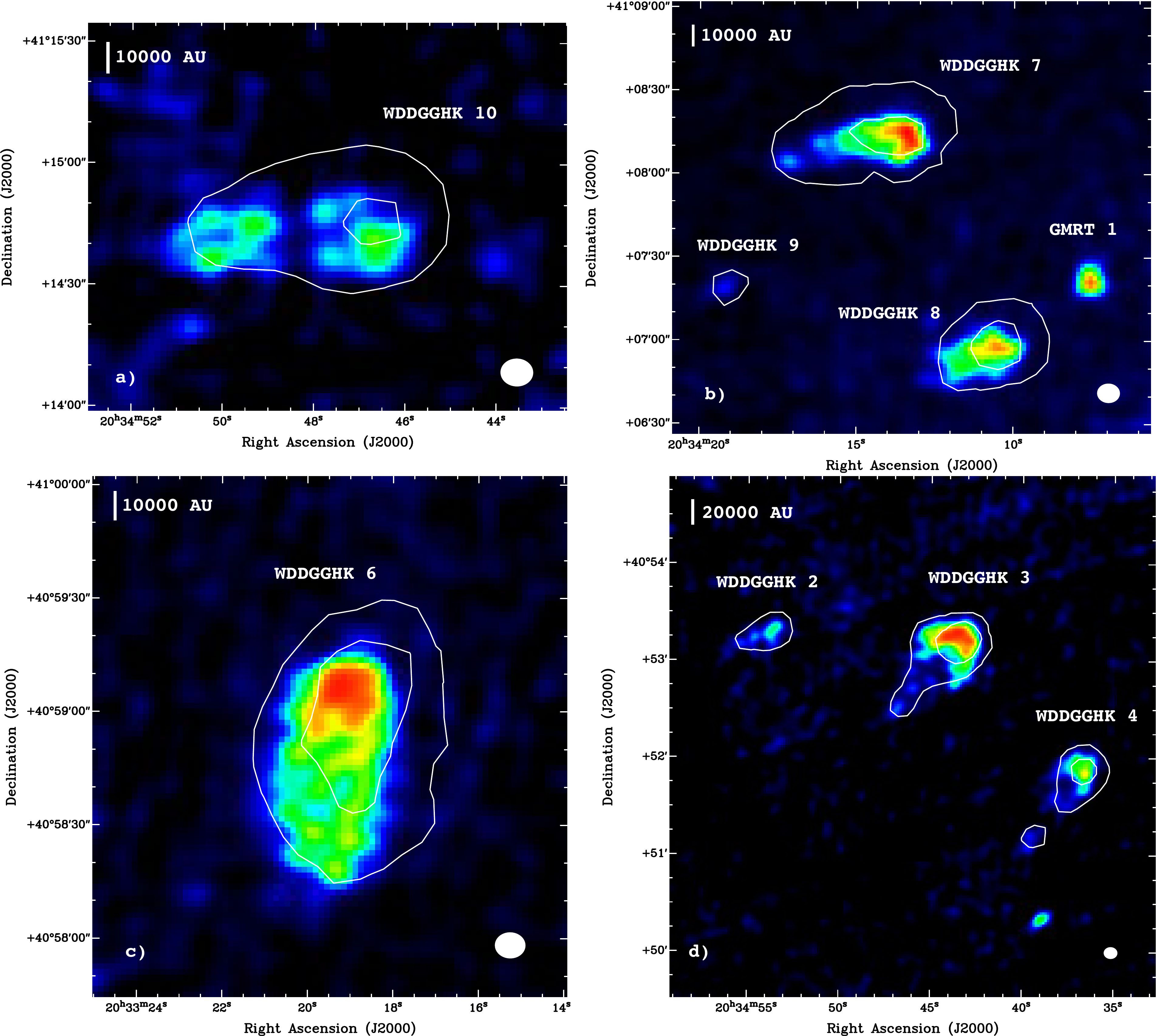}
   \caption{Zoom in to the four white boxes marked in Fig. \ref{Ffov}, 325 MHz emission in color scale, and IR contours of 470 and 1070 mJy beam$^{-1}$.}
   \label{Fzoom}%
 \end{figure*}

%-----------------------------------------------------------------

\section{Results}

\subsection{Observed parameters}
  \begin{figure*}[ht]
   \centering
   \includegraphics[width=\hsize]{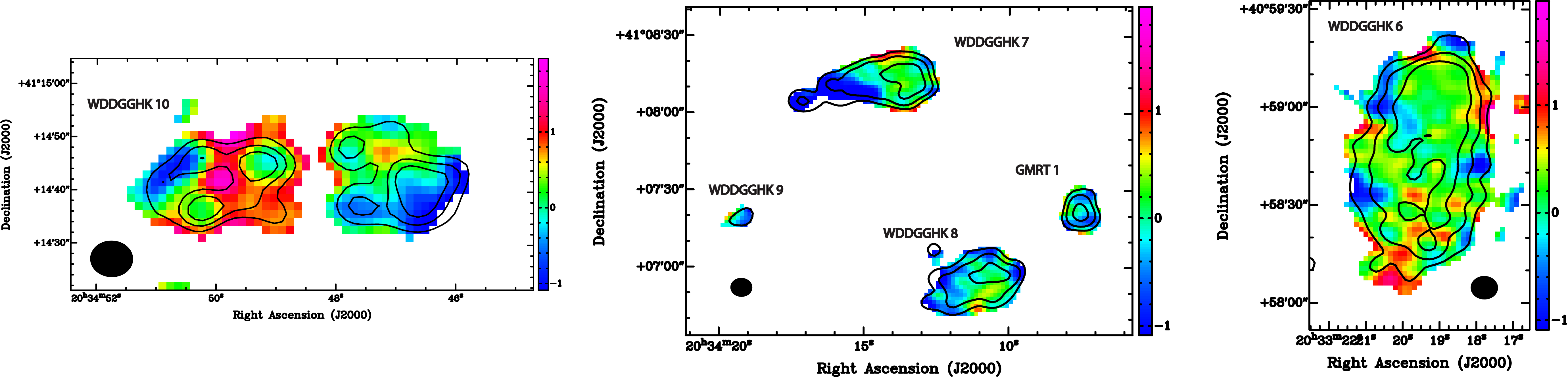}
   \caption{Spectral index distribution. {\sl Left:} source WDDGGHK 10 with 325~MHz contours of 0.5, 0.7 and 0.9~mJy~beam$^{-1}$. {\sl Center:} sources WDDGGHK 7,8 9, and GMRT 1, with 325~MHz contours of 0.5, 1 and 2~mJy~beam$^{-1}$. {\sl Right:} source WDDGGHK 6 with 325~MHz contours of 0.5, 1 and 2~mJy~beam$^{-1}$. Since the error of the spectral index is largest at the edges of the sources because the emission is weaker, extreme values of the spectral index can be generated in those areas.}
   \label{Fspix}%
  \end{figure*}
%-------------------------

The Cygnus OB2 image at 325~MHz is shown in Figure \ref{Ffov}. The primary beam of the GMRT observations includes the Cygnus OB2 association itself, prominent HII regions spreading several arcminutes, such as DR18 \citep{1999A&A...349..605C} and 080.45+01.07 \citep{2007yCat..34720557M}, radio galaxies like the double-sided NVSS J203201+413722 \citep{2006ApJ...643..238B,2007yCat..34720557M} and many others unresolved within the $\sim7\arcsec$ angular resolution, and outflows such as IRAS 20286+4105. In addition, this radio image contains most of the photoionized globules or frEGGs previously found by W12. The analysis of these frEGG sources is the focus of the present work. Only objects WDDGGHK~1 and WDDGGHK~5 from [W12]'s list lie outside of the GMRT field of view. Objects WDDGGHK~7, WDDGGHK~8 and WDDGGHK~9 were studied by \citet{2012ApJ...761L..21S} as well, who called them objects A (or Tadpole), B (or Goldfish) and C. Close to these three sources there is another radio source with negative spectral index that we named GMRT~1 and we could not classify nor discard as another frEGG.
 Following W12, the O type stars in the region are distributed over a large area and hence, although the center of the association is probably between Cyg~OB2~\#8 and Cyg~OB2~\#22 (indicated by white stars in Fig. \ref{Ffov}), other luminous stars are probably responsible for the ionization of the frEGGs. Figure \ref{Fzoom} presents four zoom maps corresponding to these locations, where the color scale indicates the flux at 325~MHz and the white contours indicate the Herschel \citep{2010A&A...518L...1P} infrared emission, very similar with the radio emission distribution, which shows the cometary morphology of the frEGGs.

Table \ref{Tsources} summarizes the positions, fluxes at 325~MHz and 610~MHz, spectral indices between both frequencies and sizes of each frEGG in the GMRT field of view. Note that sources WDDGGHK~2, WDDGGHK~3 and WDDGGHK~4 lie outside of the primary beam at 610~GHz and the corresponding fluxes are thus missing. We use a $3\sigma$ threshold to measure the integrated fluxes and estimate linear sizes, which range from about 10$\arcsec$ to 60$\arcsec$ ($15\,000$~AU to $90\,000$~AU 
at the assumed distance of 1.4~kpc), similar to those obtained from optical images (W12). The average spectral indices ($\alpha=\log(S_2/S_1) / \log{(\nu2/\nu1)}$) are mostly around zero.

We build spectral index distribution images (Fig. \ref{Fspix}) implementing in {\sc miriad} its mathematical expression. The smaller beam image was convolved with a beam corresponding to the one with the larger beam, forcing them to have the same synthesized beam. Some of these maps show differences on the spectral index across the sources, but we comment more on these in Section \ref{individual}. We determine the average optical depth ($\tau$) of the frEGGs using the following equation, considering spherical symmetry and homogeneity and the Rayleigh-Jeans approach \citep{1960TRep...59..48A},
\begin{equation}\label{eq:tau}
\tau=0.08235 (T_{\rm e}/{\rm K})^{-1.35} (\nu/{\rm GHz})^{-2.1} (E.M./{\rm pc~cm}^{-6})
\end{equation}
where $T_{\rm e}$ is the electron temperature, $\nu$ the observing frequency, and $E.M.$ the emission measure. In general, the emission is optically thin. 

Table~\ref{Tparameters} presents, besides $\tau$ values, the physical properties that we obtain analyzing the morphology of the sources (length and orientation), derived from \citet{1967ApJ...147..471M} and \citet{2016MNRAS.459.1248A} expressions for HII regions, evaluated at the frequency of 325~MHz. In doing this, as a first order approximation we consider a model of free-free emission in an optically thin homogeneous sphere. We adopted the values $T_{\rm e} = 10^4$~K and $d = 1.4$~kpc \citep{2012A&A...539A..79R} to estimate the electron density, mass of ionized gas, emission measure and rate of Lymann photons needed to ionize each source, the latter with $(N_{\rm i}/s^{-1})=7.6x10^{43} (S_{\nu}/\mathrm{mJy})  (\nu/\mathrm{GHz})^{0.1}  (T_{\rm e}/10^4{\rm K})^{-0.45} (D/\mathrm{kpc})^2$ \citep{1987ApJ...314..535G}.
The densities are typical of photoionized regions ($10^2$~cm$^{-3}$) with ionized gas masses of 0.002~M$_{\odot}$--0.2~M$_{\odot}$. The photon rates range between 10$^{43}$~s$^{-1}$--10$^{45}$~s$^{-1}$, which can be produced by early B-type stars. We give more insight on the physical properties in Section \ref{individual}, where we review the main characteristics of each frEGG in our sample.

\begin{table*}
\caption{Observed parameters for the frEGGs on Cygnus OB2.}     % title of Table
\label{Tsources}      % is used to refer this table in the text
\centering                          % used for centering table
\begin{tabular}{l c c c c c c c c c}        % centered columns (4 columns)
\hline\hline                 % inserts double horizontal lines
 ID & RA(J2000) & Dec(J2000) & $S_{325}$& $S_{610}$ & $\alpha$ & $\theta_{325}$ & $P.A._{325}$ & $\theta_{610}$ & $P.A._{610}$ \\  % table heading
 & (h,m,s)  & ($\degr$,$\arcmin$,$\arcsec$) & (mJy) & (mJy) & & ($\arcsec$)& ($\degr$) & ($\arcsec$) & ($\degr$) \\
 \hline                        % inserts single horizontal line
WDDGGHK2 & 20:34:53.55 & +40:53:20.34 & 5.5$\pm$0.4  & ---& --- & 27 & 132 & --- & ---\\ % inserting body of the table
WDDGGHK3 &20:34:43.49 & +40:53:17.98 & 54$\pm$1 & --- & --- & 57 & 138 & --- & ---\\
WDDGGHK4 &20:34:36.577 & +40:51:49.90 & 15.2$\pm$0.6 & ---& --- & 36 & 141 & --- & --- \\
WDDGGHK6 & 20:33:19.276 & +40:59:06.26 & 68$\pm$1 & 79$\pm$1 & 0.22$\pm$0.07 & 35 & 177 & 33 & 176 \\
WDDGGHK7 & 20:34:13.385 & +41:08:15.11 & 19.8$\pm$0.6 & 17.8$\pm$0.8 & -0.2$\pm$0.2 & 28 & 108 & 25 & 98\\
WDDGGHK8 & 20:34:10.438 & +41:06:57.24 & 15.2$\pm$0.4 & 13.7$\pm$0.6 & -0.2$\pm$0.2 & 21 & 112 & 22 & 118\\
%GMRT1 & 20:34:07.525 & +41:07:21.37 & 4.2$\pm$0.6 & 2.63 & 3.6$\pm$0.5 & 2.43 & -0.244$\pm$0.728 & 12 & 90\\
WDDGGHK9 & 20:34:19.207 & +41:07:19.34 & 0.4$\pm$0.1 & 0.4$\pm$0.1 & 0$\pm$1 & 11 & 127 & 8 & 119\\
WDDGGHK10 & 20:34:46.639 & +41:14:41.82 & 9.9$\pm$0.2 & 12.2$\pm$0.3 & 0.3$\pm$0.1 & 32 & 90 & 28 & 89\\ 
\hline                                   %inserts single line
\end{tabular}
\tablefoot{(1) Geometric sizes, $\theta$, obtained from $\sqrt{ab}$, where $a$ and $b$ are major and minor axis respectively. (2) Typical uncertainties in the measured sizes are 1.5$\arcsec$, and in $P.A.$ are 8$\degr$.}
\end{table*}
\begin{table*}
\caption{Physical properties for the frEGGs of Cygnus OB2.}     % title of Table
\label{Tparameters}      % is used to refer this table in the text
\centering                          % used for centering table
\begin{tabular}{l c c c c c c c }        % centered columns (4 columns)
\hline\hline                 % inserts double horizontal lines
 ID & $n_e$ & $M_{\rm ion}$ & $E.M.$ & $\tau$ & $N_{\rm Lyman}$  \\  % table heading
    & (cm$^{-3})$  & (M$_{\odot})$ &(pc~cm$^{-6})$& & $ (\rm s^{-1})$  \\
\hline                        % inserts single horizontal line
WDDGGHK2 & 100$\pm$10 & 0.025$\pm$0.003 & 2500$\pm$700 & 0.009$\pm$0.003 & $7.32\times10^{44}$ \\      % inserting body of the table
WDDGGHK3 & 98$\pm$5 & 0.24$\pm$0.01 & 55500$\pm$700 & 0.019$\pm$0.002 & $7.18\times10^{45}$   \\
WDDGGHK4 & 103$\pm$8 & 0.063$\pm$0.005 & 3800$\pm$800 & 0.013$\pm$0.003 & $1.99\times10^{45}$  \\
WDDGGHK6 & 230$\pm$20 & 0.13$\pm$0.09 & 18000$\pm$3000 & 0.06$\pm$0.01 & $9.05\times10^{45}$  \\
WDDGGHK7 & 170$\pm$20 & 0.050$\pm$0.005 & 8000$\pm$2000 & 0.029$\pm$0.007 & $2.66\times10^{45}$  \\
WDDGGHK8 & 230$\pm$30 & 0.03$\pm$0.03 & 11000$\pm$ 3000 & 0.04$\pm$0.01 & $1.99\times10^{45}$  \\
%GMRT1 & 284$\pm$73 & 0.006$\pm$0.002 & 9590$\pm$6170 & 0.050$\pm$0.003 & 45.1 \\
WDDGGHK9 & 100$\pm$30 & 0.002$\pm$0.001 & 1100$\pm$800 & 0.004$\pm$0.003 & $5.22\times10^{43}$  \\
WDDGGHK10 & 100$\pm$8 & 0.042$\pm$0.003 & 3200$\pm$600 & 0.011$\pm$0.002 & $1.33\times10^{45}$  \\ 
\hline                                   %inserts single line
\end{tabular}
\tablefoot{The values of $n_e$, $M_{\rm ion}$, $E.M.$ were determined with equations (5), (6) and (7) from \citealt{2016MNRAS.459.1248A}.}
\end{table*}

\subsection{Individual sources}
\label{individual}
The observed frEGGs are distributed south of the OB2 association and are grouped in four distinct regions (boxes \textit{a}, \textit{b}, \textit{c} and \textit{d} in Fig.~\ref{Ffov}). There are two isolated objects (WDDGGHK~6 and WDDGGHK~10) and two groups of three objects each (boxes \textit{b} and \textit{d}, respectively in Figs. \ref{Ffov} and \ref{Fzoom}). All the sources show an elongated morphology and spectral indices around zero. The morphology in radio wavelengths matches perfectly the emission detected by Herschel at 70\um. The electron density and the orientation of the sources is similar within the members of the two mentioned groups. We summarize the main findings of each object below.

\textbf{WDDGGHK~2}: This elongated (37000~AU) source, also known as IPHASX J2034533+405321 is located at the southeast edge of the 325~MHz primary beam (box \textit{d} in Figs. \ref{Ffov} and \ref{Fzoom}), lying outside of the 610~MHz primary beam. It is grouped with objects WDDGGHK~3 and WDDGGHK~4, with which shares a northwest-southeast orientation and a similar electron density around 90-100\cc. As with these other two objects, the tail of WDDGGHK~2 seems to comprise two lateral threads. It has been also detected in infrared, submillimeter and optical wavelengths \citep{2012ApJ...746L..21W,2016A&A...591A..40S}. 

\textbf{WDDGGHK~3}: It is located about 0.5~pc west of WDDGGHK~2 and 0.5~pc east of WDDGGHK~4, and is one of the strongest sources of our sample at 325~MHz. This object presents a clear cometary shape with a prominent head, showing a bowshock structure (also detected in optical wavelengths, see \citealt{2002A&A...389..874C}) and a faint tail due southeast. Its 80\,000~AU length makes it the largest frEGG in our sample. Additionally, it has the largest content of ionized gas (0.236 M$_\sun$), needing an ionizing photon rate of 10$^{45.8}\rm s^{-1}$. WDDGGHK~3 contains a protostar probably associated with IRAS~20328+4042, a 100 L$_\sun$ young stellar object embedded in a massive gaseous envelope or disk \citep{2002A&A...389..874C,2011ApJ...727..114R}.

\textbf{WDDGGHK~4}: This elongated source is spatially resolved at 325~MHz (50\,000~AU in length). It splits into two clear parts: a head with a bowshock shape pointing northwest ($P.A.=141\degr$) and a frayed tail, comprised of faint {shreds}, one of them disconnected from the rest (Fig. \ref{Fzoom}). Previous FIR observations between 500\um and 60\um toward the head of this object have identified a dusty core, BLAST~C62, possibly harboring a stellar nursery \citep{2011ApJ...727..114R}. About $1.8'$ south of WDDGGHK~4's head there is an unresolved source with a flux density of 2.7 mJy (at 325 MHz) that we do not include in our analysis of frEGGs, although it could be related with this group of photoionized sources because of its proximity and its similar orientation,  { maybe a tail emanating behind the bow shock.}

\textbf{WDDGGHK~6}: Box \textit{c} in Figure \ref{Ffov} contains this isolated source (see also, Figs. \ref{Fzoom} and \ref{Fspix}) with a north-south orientation ($P.A.=177\degr$). As many of these objects, it has a very strong arc-shaped head, followed by a frayed tail with two or three shreds, also observed in H$\alpha$ and 8\um images (W12). 
%The shreds are curved, possibly wiggling and twisting; this behavior is also seen in the optical images. In addition, there is an asymmetry in the brightness of the WDDGGHK~6 shreds, being stronger the radio emission from those to the west. Its opacity is the greatest in our frEEGs sample, and 
It is one of the densest sources as well (230\cc)\footnote{This value represents a lower limit.} It needs an ionization photon rate of 10$^{45.9}~{\rm s^{-1}}$. \citet{2011ApJ...727..114R} associated this photoionized region with IRAS20315+4046, whose emission may be originated by a 400 L$_\sun$ protostar(s).

\textbf{WDDGGHK~7}: Dubbed the \textit{Tadpole} because of its morphological resemblance with the aquatic animal, it has been the focus of several works \citep{2012ApJ...746L..21W, 2012ApJ...761L..21S,2014ApJ...793...56G,2016A&A...591A..40S}. It is $\sim12\arcmin$ southwest of DR18 (box \textit{b} in Fig.~\ref{Ffov}) and it consists of the characteristic bowshock head, followed by a fainter tail (Figs. \ref{Fzoom} and \ref{Fspix}). The object is oriented almost east-west ($P.A.=108\degr$), similarly to its frEGGs neighbors WDDGGHK~8 and  {possibly} WDDGGHK~9. From the radio emission detected with the GMRT we derive an average spectral index of $-0.2 \pm 0.2$. However, the spatial distribution of the spectral index is not homogeneous, varying from positive at the head to negative at the tail. In section \ref{tadp} we analyze its spectral energy distribution (SED) in detail.

\textbf{WDDGGHK~8}: This source, also called the \textit{Goldfish}, is about $1\arcmin$ south of the Tadpole. It is as dense as the Tadpole (233\cc)
but with a slightly smaller ionized mass. We find an average spectral index of $-0.16 \pm 0.19$ which, in this case, is homogeneously distributed throughout the source (i.e., no apparent variations), that within the errors it is consistent with emission from an optically thin plasma.

\textbf{WDDGGHK~9}: Another neighbor of the Tadpole but barely detected by GMRT observations. The weakness of the emission makes the estimate of its physical properties very uncertain. We estimate its length to be $\sim15\,000$~AU and its orientation to be $P.A.=127\degr$, similar to those of the Tadpole and the Goldfish. 

\textbf{WDDGGHK~10}: Isolated source located inside the box \textit{a} in Figure \ref{Ffov}. It is $3\farcm5$ west of the prominent $4\arcmin$ long bowshock of the DR18 ionized region. Figure \ref{Fzoom} shows that WDDGGHK~10 comprises two main radio components, oriented east-west, interestingly the same orientation aimed by the DR18 bow shock. NIR and optical images (W12) show WDDGGHK~10 has a head-tail bow-shock morphology, in good agreement with the GMRT image, which displays stronger emission toward the head of the shock (west) and somewhat fainter toward the tail (east). At radio wavelengths, the contrast between head and tail is not as large as in the other frEGGs studied here. Maybe a reason for this feature could be that the shreds of the tail of this object are apparently twisted and cross each other at the position of the eastern radio peak (noticeable in the optical and NIR images; \citealt{2012ApJ...746L..21W}). This would explain the emission enhancement from the tail. On average, this source has the steepest spectral index in our sample ($\alpha=0.3$). It is worth to note that the two isolated frEGGs have average positive spectral indices, while the other sources have flat or slightly negative spectral indices. The spatial distribution of the spectral index toward WDDGGHK~10  {(Fig.~\ref{Fspix}}) 
is also interesting since it varies from the head at $\alpha\sim-0.1$, growing in the center to $\alpha\sim0.5$ and decreasing toward the easternmost part of the tail to $\alpha\sim0.3$. 
These variations in the spectral index could be due to differences in the optical depth along the source.

Regarding the protostellar content of WDDGGHK~10, the only IR source inside the photoionized region found by \citet{2012ApJ...746L..21W} is IRAS~20329+4104, which is about $10\arcsec$ (14\,000~AU) away from the peak emission in radio from this frEGG.

\subsection{Tadpole spectral energy distribution}\label{tadp}

In this section we analyze the SED of the object WDDGGHK7, the Tadpole, built up with data from previous radio and infrared observations (Table \ref{Tsed}). We selected the Tadpole as an example of a frEGG with a flat/negative spectral index (regarding to our data), and the one with a suspicious negative spectral index at higher frequencies \citep{2012ApJ...761L..21S}.

\begin{table}
\caption{Flux density values of WDDGGHK7 (the Tadpole) from radio to the IR range.} % title of Table
%---------> 

\label{Tsed}      % is used to refer this table in the text
\centering                          % used for centering table
\begin{tabular}{l c c c c}        % centered columns (4 columns)
\hline\hline                 % inserts double horizontal lines
 $\nu$ & $S_{\nu}$ & Resolution & Reference \\  % table heading
  (GHz)  & (Jy) & ($\arcsec$)  \\
 \hline                        % inserts single horizontal line
0.325 & 0.019$\pm$0.0006 & 7.8 & This work \\ % inserting body of the table
0.350 & 0.05$\pm$0.006 & 55 & S. Gunawan et al. 2003 \\
0.610 & 0.017$\pm$0.0008 & 7.6 & This work \\
1.4 & 0.0084$\pm$0.0005 & 13 & S. Gunawan et al. 2003 \\
1.4C & 0.044$\pm$0.007 & 55 & S. Gunawan et al. 2003 \\
8.5 & 0.055$\pm$0.0012 & 3.2 & Sahai et al. 2012\\ 
22.5 & 0.030$\pm$0.0014 & 3.2 & Sahai et al. 2012 \\
%600 & 57.6$\pm$3.5 & 60 & Roy et al. 2011 \\
%857 & 94.2$\pm$3.4 & 42 & Roy et al. 2011 \\
%1200 & 168.6$\pm$7.3 & 30 & Roy et al. 2011 \\
%3000 & 70.8 & 120 & Roy et al. 2011 \\
%5000 & 76.6 & 120 & Roy et al. 2011 \\
\hline                                   %inserts single line
\end{tabular}
\tablefoot {1.4C: Convolved to the 350 MHz beam.}
\end{table}

\begin{figure}
   \centering
 \includegraphics[width=0.29\textwidth, angle=270]{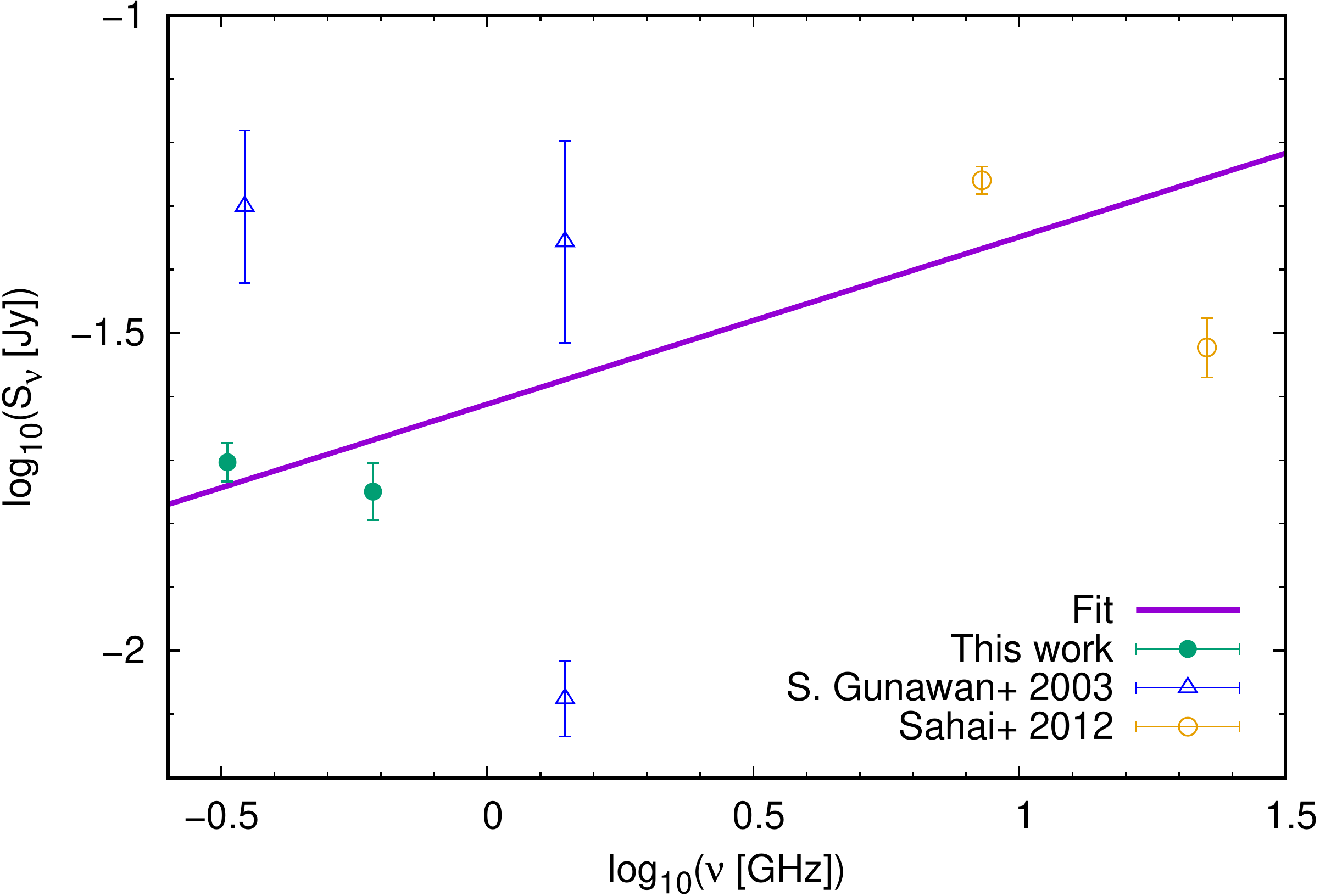}
   \caption{SED of the Tadpole (WDDGGHK7). We have fitted a power-law to the available radio interferometer data (dashed green line); the purple line shows the sum of both components.}
    \label{Fsed}
\end{figure}

%In order to fit the SED  at IR frequencies, we use a modified Planck law $S_\nu \propto \nu^\beta \, B_\nu$, where $B_\nu$ is the Planck function and $\beta$ is the opacity spectral energy of the dust. The value of $\beta$ has to be above 0.5 in order not to overestimate the emission at 22~GHz reported by \citet{2012ApJ...761L..21S}. We fix $\beta = 1$ and obtain a temperature $T \approx 26$~K for the dust.

At radio wavelengths we use a power-law obtaining an average slope of 0.26 (see Fig.~\ref{Fsed}) which is consistent with thermal emission from a partially optically thin plasma. The thermal emission is expected to be stationary unless the thermodynamical quantities of the gas (e.g., $n_\mathrm{e}$) suffer a quick evolution in a year--scale, which is highly unlikely. Therefore, the mismatch between our observations and the values reported by \citet{2003ApJS..149..123S} should be explained by differences in beamsize and calibration errors. 

Although partial power-law fits in specific ranges of the radio spectrum show negative spectral indices, the overall index from frequencies below 1~GHz to 22.5~GHz is consistent with thermal emission. If the data at 8 and 22 GHz by \citet{2012ApJ...761L..21S} is interpreted as non-thermal emission, one would expect that the flux detected at lower frequencies of 0.325-1.4 GHz to be much higher. The only way to reconcile this would be to summon the action of absorption processes which reduce the observed low-frequency emission \citep{Melrose1980}. However, this would lead to a positive spectral index at low frequencies which is in contradiction to the flat spectrum reported we observe. Therefore, we favour an explanation involving calibration uncertainties and/or the presence of diffuse emission which is lost (not detected) at high frequencies (as the LAS is smaller at higher frequencies), which could artificially generate the observed slightly negative spectral index.  More observations at intermediate frequencies will help to setle this issue.

\section{Discussion}

In this section, we present a short discussion about the sources causing the ionization of the proplyd-like objects found in Cygnus~OB2. We also discuss the possible origin of the flat radio-spectral indices of some of them, focusing particularly on the Tadpole, which has been observed at several wavelengths and its radio emission has been interpreted as non-thermal in the past \citep{2012ApJ...761L..21S}.  

\subsection{Sources causing the ionization}

\begin{table*}
\caption{Differences between needed ionizing photon rates and those expected from nearby massive Cygnus OB2 stars.} % title of Table
%---------> 
\label{Tnlyman}      % is used to refer this table in the text
\centering                          % used for centering table
%\begin{tabular}{l c c c c c c c c c}        % centered columns (4 columns)
\begin{tabular}{l@{~~~}c@{~~~}c@{~~~}c@{~~~}c@{~~~}c@{~~~}c@{~~~}c@{~~~}c@{~~~}c}
\hline\hline                 % inserts double horizontal lines
 Source & $N_{\rm i}^{\rm needed}$ & $N^\prime_{\rm i}$ \#9 & $\Delta N_{\rm i}$ \#9 & $N^\prime_{\rm i}$ \#22 & $\Delta N_{\rm i}$ \#22 & $N^\prime_{\rm i}$ \#8 & $\Delta N_{\rm i}$ \#8 & $N^\prime_{\rm i}$ MT~516 & $\Delta N_{\rm i}$ MT~516  \\  % table heading
   & (s$^{-1}$) & (s$^{-1}$) & (\%) & (s$^{-1}$) & (\%) & (s$^{-1}$) & (\%) & (s$^{-1}$) & (\%)   \\
  (1) & (2) & (3) & (4) & (5) & (6) & (7) & (8) & (9) & (10)   \\
 \hline                        % inserts single horizontal line
WDDGGHK2 & $7.32\times10^{44}$ & $6.19\times10^{44}$ & -15  & $5.02\times10^{44}$ & -31   & $4.30\times10^{44}$ & -41  & $1.31\times10^{44}$ & -82  \\
WDDGGHK3 & $7.19\times10^{45}$ & $4.25\times10^{45}$ & -41  & $3.47\times10^{45}$ & -52   & $2.90\times10^{45}$ & -60  & $9.35\times10^{44}$ & -87  \\
WDDGGHK4 & $2.02\times10^{45}$ & $2.21\times10^{45}$ & 9    & $1.81\times10^{45}$ & -10   & $1.49\times10^{45}$ & -26  & $4.88\times10^{44}$ & -76  \\
WDDGGHK6 & $9.05\times10^{45}$ & $8.35\times10^{45}$ & -8   & $7.93\times10^{45}$ & -12   & $4.48\times10^{45}$ & -50  & $2.85\times10^{45}$ & -68  \\
WDDGGHK7 & $2.63\times10^{45}$ & $4.16\times10^{45}$ & 58   & $3.37\times10^{45}$ & 28    & $2.69\times10^{45}$ & 2    & $1.18\times10^{45}$ & -55  \\
WDDGGHK8 & $2.02\times10^{45}$ & $3.71\times10^{45}$ & 84   & $3.07\times10^{45}$ & 52    & $2.33\times10^{45}$ & 15   & $1.16\times10^{45}$ & -42  \\
WDDGGHK9 & $5.32\times10^{43}$ & $9.43\times10^{44}$ & 1671 & $7.66\times10^{44}$ & 1338  & $6.26\times10^{44}$ & 1075 & $2.54\times10^{44}$ & 377  \\
WDDGGHK10 & $1.31\times10^{45}$ & $3.37\times10^{45}$ & 157 & $2.44\times10^{45}$ & 86    & $2.85\times10^{45}$ & 118  & $5.39\times10^{44}$ & -59  \\
\hline                                   %inserts single line
\end{tabular}
\tablefoot{The distances between the ionizing candidates and the WDDGGHK~3 group range 1300$\arcsec$-1900$\arcsec$ ($\sim13$~pc); to WDDGGHK~6, distances range 600$\arcsec$-1200$\arcsec$ ($\sim7.5$~pc); to the WDDGGHK~7 group, distances range 550$\arcsec$-900$\arcsec$ ($\sim6$~pc); to WDDGGHK~10, distances range $1000\arcsec$-1100$\arcsec$ ($\sim8.5$~pc). (1): Name of the source; (2): Rate of ionizing Lyman photons needed to produce the measured radio emission; (3), (5), (7) and (9): Rates of Lyman photons incident at each of the frEGGs (see text) and produced by Cygnus~OB2 stars \#9 (O5~I and O3.5~III), \#22 (O5~I and O3.5~III), \#8 (considering \#8A and \#8B, which comprises an O6~I plus an O5.5~III, and an O6.5~III), and the star RPL841 (O5.5~V), respectively. The spectral type of the stars were adopted from \citet{2015yCat..74490741W}. The ionizing photon rates from the corresponding spectral type, using the models based on observational $T_{eff}$ scales by \citet{2005A&A...436.1049M}; (4), (6), (8) and (10): Difference in \% of the needed and the provided ionizing photon rates, $100 \times (N^\prime_{\rm i}-N_{\rm i}^{\rm needed})/N_{\rm i}^{\rm needed}$. Negative percentages indicate a deficit of ionizing photons from the star, while positive percentages indicate the opposite, an excess of ionizing photons (or a larger than projected distance to the object).}
\end{table*}

\begin{table*}
\caption{Differences between the $P.A.$ of frEGGs and its orientation with respect to nearby massive Cygnus OB2 stars.} % title of Table
%---------> 
\label{Torientation}      % is used to refer this table in the text
\centering                          % used for centering table
%\begin{tabular}{l c c c c c c c c c}        % centered columns (4 columns)
\begin{tabular}{l@{~~~}c@{~~~}c@{~~~}c@{~~~}c@{~~~}c@{~~~}c@{~~~}c@{~~~}c@{~~~}c@{~~~}c}
\hline\hline                 % inserts double horizontal lines
 Source & $P.A.$ & $P.A.$ from \#9 & $\Delta$ \#9 & $P.A.$ from \#22 & $\Delta$ \#22 & $P.A.$ from \#8 & $\Delta$ \#8 & $P.A.$ from MT~516 & $\Delta$ MT~516  \\  % table heading
   & ($\degr$) & ($\degr$) & ($\sigma$) & ($\degr$) & ($\sigma$) & ($\degr$) & ($\sigma$) & ($\degr$) & ($\sigma$)   \\
  (1) & (2) & (3) & (4) & (5) & (6) & (7) & (8) & (9) & (10)   \\
 \hline                        % inserts single horizontal line
WDDGGHK2  & 132 & 138 & 0.8 & 136 & 0.5 & 145 & 1.7 & 139 & 0.9 \\
WDDGGHK3  & 138 & 141 & 0.4 & 139 & 0.1 & 147 & 1.2 & 143 & 0.7 \\
WDDGGHK4  & 141 & 144 & 0.4 & 142 & 0.1 & 151 & 1.3 & 147 & 0.8 \\
WDDGGHK6  & 177 & 180 & 0.4 & 172 & 0.7 & 178 & 0.1 & 184 & 0.9 \\
WDDGGHK7  & 108 & 120 & 1.6 & 116 & 1.1 & 136 & 3.7 & 118 & 1.3 \\
WDDGGHK8  & 112 & 125 & 1.7 & 121 & 1.2 & 141 & 3.9 & 125 & 1.7 \\
WDDGGHK9  & 127 & 121 & 0.8 & 116 & 1.5 & 135 & 1.1 & 119 & 1.1 \\
WDDGGHK10 &  90 &  91 & 0.1 &  88 & 0.3 & 105 & 2.0 &  85 & 0.7 \\
\hline                                   %inserts single line
\end{tabular}
\tablefoot {(1): Name of the source; (2): Position angle ($P.A.$) of each frEGG (we estimate the typical uncertainty of these measurements to be $8\degr$); (3), (5), (7) and (9): $P.A.$ measured from the considered Cygnus~OB2 ionizing candidates to each frEGG; (4), (6), (8) and (10): Difference (absolute value) between the frEGG $P.A.s$ and the orientation with respect to the ionizing star candidates, in terms of the uncertainty of the frEGG's $P.A.$ measurement.}
\end{table*}

In Table \ref{Tnlyman} we report the photon rate necessary to ionize the Cygnus~OB2 frEGGs. It ranges between $5~\times~10^{43}$ and almost $10^{46}\rm s^{-1}$, corresponding to main sequence stars with spectral types B3-B1 \citep{1973AJ.....78..929P}. According to \citet{2012ApJ...746L..21W}, the ionizing stars responsible for photoionizing all the Cygnus~OB2 frEGGs could be OB2~\#8 ($RA$=20:33:15.07, $Dec$=41:18:50.47) or OB2~\#22 ($RA$=20:33:08.79, $Dec$=41:13:18.21), although they do not provide quantitative evidence. These systems (see e.g., \citealt{2015yCat..74490741W}) are a binary comprised by an O6~I plus an O4.5~III stars (OB2~\#8A) along with an O6.5~III (OB2~\#8B), and another binary with an O3~If plus an O6~V stars (OB2~\#22). The agregated ionization rates of these systems are $7.52\times10^{49}$ and $7.00\times10^{49}$ photons per second, respectively. We use the Lyman photon rates tabulated by \citet{2005A&A...436.1049M}, which have into account line-blanketing and wind effects. In addition to these stellar systems we have tested another two massive systems favorably located in the area so that they could also be responsible for ionizing all the frEGGs \citep{2015yCat..74490741W}: OB2~\#9 (comprised by an O5~I plus an O3.5~III), and MT~516 (an O5.5~V star, with the number 516 in the system by \citealt{1991AJ....101.1408M}). These spectral types lead to total ionizing photon fluxes of $9.30\times10^{49}$ and $1.26\times10^{49}$ per second, respectively. From all four massive stellar systems, the only with accurate Gaia~DR2 parallax reported is OB2~\#9 ($\pi=0.60$), at a distance of 1663~pc.   
Now, to test if these stars are ionizing the Cygnus~OB2 frEGGs we use as a first criterion the prescription in \citet{1987ApJ...314..535G}. In particular, their equation 4, 
$$N_\mathrm{i}=N_\mathrm{i0}\frac{\Omega}{4\pi}e^{-\tau_{\rm Lyman}}\quad,$$
provides the ionizing photon rate $N_\mathrm{i}$ incident on a source that subtends a solid angle $\Omega$ as seen from a star with a $N_\mathrm{i0}$ ionizing photon rate. Here $\Omega=\pi*(r/d)^2$, where $d$ is the distance from the ionized object to the ionizing source and $r$ the projected radius of the object as seen from the ionizing stars. For the calculations we use the projected distance as $d$ (therefore the obtained $N_\mathrm{i}$ would be upper limits), and assume that the Lyman continuum optical depth $\tau_{\rm Lyman}$ is negligible.  We compare these rates with the ionization photon flux needed to ionize the frEGGs derived from their associated radio continuum emission (Table \ref{Tnlyman}). A second criterion to evaluate if a star is effectively ionizing the OB2 frEGGs is the relative agreement between the orientation of the projected line linking the star with the head of a frEGG and the position angle of that frEGG (Table \ref{Torientation}).
The positional alignment is quite good for stars \#22, \#9 and MT~516, with deviations less than 1.5 times the $P.A.$ uncertainty, in general (although \#9 and WDDGGHK~7-8 are not that well aligned). On the contrary, \#8 shows large deviations for some frEGGs, whether in the group of WDDGGHK~2, WDDGGHK~7 or with WDDGGHK~10. Thus, system OB2~\#8 seems not adequately located to be a good candidate. 
Regarding the Lyman photon fluxes, none of the stellar systems provide a high enough rate to ionize the southern WDDGGHK~2-4 frEGGs, specially WDDGGHK~3, which needs $7.2\times10^{45}$ photons~s$^{-1}$. The single star MT~516, although is closer to all the frEGGs is of a later type and cannot ionize the sources on its own. Only systems \#9 and \#22 are close to the required values. They provide higher than needed fluxes for frEGGs WDDGGHK~7-10, slightly lower for WDDGGHK~6 and half of the needed rate for WDDGGHK~3. The reason for a higher than needed rate may be simply a projection effect, while lower than needed rates may indicate that a combination of fluxes from different stellar systems are responsible for the final ionization. When adding the contribution from both stellar systems (\#9 and \#22), the resulting photon flux is higher than needed, which may be lowered if unprojected distances are used in the calculations. Therefore, the most favorable ionizing sources for the Cygnus~OB2 frEGGs are OB2~\#9 and OB2~\#22, and small amounts of photons would also be provided by the other stellar systems. It is noticeable, that all the stellar candidates could provide Lyman photon fluxes hundreds of times the rate needed to ionize WDDGGHK~9. 
%This could indicate that this frEGG is actually about 30~pc away from the others (although it is very close in projection), or that all its material is ionized with a smaller amount of photons than those provided by the stars considered.     

\subsection{Spectral indices}
%We next deal with the radio spectral indices of the observed frEGGs. 
In Table \ref{Tsources} we report the measured spectral indices extracted from the fluxes obtained at the two observing frequencies: 325~MHz and 610~MHz. The determination of the spectral indices produces small values with relatively large uncertainties. All spectral indices are below the canonical 0.6 index typical of HII regions (e.g., \citealt{1975A&A....38..451C}). Moreover, two of the frEGGs (WDDGGHK~7 and WDDGGHK~8) show negative spectral indices of $-0.2\pm0.2$, consistent with flat indices if the uncertainties are taken into account. As already pointed out in previous sections, a negative spectral index may indicate the presence of non-thermal emission. However, in this case, the reason behind the low spectral indices measured in the Cygnus~OB2 frEGGs may be the existence of significant amounts of diffuse gas (extended and tenuous emission), preferentially detected at lower frequencies by the GMRT (see also \citealt{2017MNRAS.465.4753R}). Supporting this argument, we check for 
any well-known HII region lying within the field of view of our Cygnus~OB2 GMRT observations. We found for instance DR~18.
Using our observations, we obtain a radio-spectral index of 0.3$\pm$0.01 for this extended object, which is also lower than the expected 0.6 value. For DR~18, it is evident also that for extended objects, the GMRT is more sensitive to the extended weak emission at 325~MHz than at 610~MHz. This effect can be quantitatively accounted for by reporting the number of pixels with significant emission at each frequency. Again, the continuum emission of DR~18 at 325~MHz spreads on 4\% more pixels than at 610~MHz. Similar results are obtained for most of the frEGGs: for WDDGGHK~8 the diffuse emission extends over a 10\% more pixels at 325~MHz, while for WDDGGHK~10, this percentage rises to 30\%. This effect is more pronounced in the tail of WDDGGHK~7 and WDDGGHK~8, which is the region with more diffuse emission and negative values of the spectral index (Fig. \ref{Fspix}). As a consequence the radio-spectral indices are smaller than they are in reality and therefore we are no further interpreting them as originated by non-thermal emission. We just make the caveat for future usage of our results that diffuse emission have to be accounted for when estimating the spectral index in these sources.  
Then the values of $\tau$, $n_e$, $M_{\rm ion}$ and $E.M.$ from Table~2 were determined at the frequency of 325~MHz since they are not affected by the loss of emission at the frequency of 610~MHz.

A special case among the studied frEGGs is that WDDGGHK~7, the Tadpole, since \citet{2012ApJ...761L..21S} have reported the detection of a negative spectral index between 8.5~GHz and 22.0~GHz. Our low-frequency spectral index is consistent with a slightly negative or flat value, which cannot be reconciled with \citet{2012ApJ...761L..21S}'s data (Figure \ref{Fsed}). As already mentioned, we only detect negative spectral index values between $-1.0$ and $-0.5$ along the tail of this object (Fig. \ref{Fspix}). Hence, in case there is any non-thermal emission it should belong to this part of the object. However, 15\% of Tadpole's pixels at 325~MHz are not seen at 610~MHz and most of them are lost in the tail. This suggests that Tadpole's tail is probably affected by diffuse emission, which artificially produces the negative index. When studying together the GMRT measured fluxes with those of \citet{2012ApJ...761L..21S} for the Tadpole (Figure \ref{Fsed}) a power-law fit produces a slope of $0.3\pm0.1$. This is probably a more reliable value for the radio-spectral index of this source. It shows that thermal emission dominates Tadpole's radio emission (a black-body is a good fit for its infrared emission in addition), although more data at intermediate and larger frequencies would improve this fit to finally constrain the nature of the radio emission.

\section{Summary and conclusions}

We study the externally ionized objects of the Cygnus\,OB2 region previously identified in \citet{2012ApJ...746L..21W} through GMRT 325~MHz and 610~MHz observations. After describing them observationally we determine their physical properties, such as their electronic densities, ionized mass, optical depth and the amount of photons needed to be ionized. We compare the ionizing photon rate with that produced by different massive stars in the neighborhood searching for good ionizing candidates. We also compare the orientation of the frEGGs with the position angle between the ionizing stars and the frEGGs. We conclude that the stellar systems Cyg~OB2~\#9 and Cyg~OB2~\#22 are probably responsible for their ionization.
 
We also obtain radio spectral indices between 325~MHz and 610~MHz for five frEGGs of our sample. Some sources show an average flat spectral index consistent with thermal free-free emission. However, some of the spectral index maps show regions of negative values, that we interpret as a possible effect of the presence of diffuse gas, since the interferometer would be more sensitive to the extended diffuse gas at low frequencies, resulting in spectral index values close or below to zero. Still our observations are not conclusive about the existence of regions with non-thermal emission in the frEGGs, although they suggest that the main contribution to the continuum radio-emission is thermal.
 
Regarding the object WDDGGHK~7 (the Tadpole), thought to be non-thermal at radio-wavelengths, we collect data in the literature to build its SED. We fit a power-law function to the radio-continuum measurements obtaining a 0.3 spectral index consistent with thermal emission. Hence, according to our data, most of the radio-emission from frEGGs would not be non-thermal.

\begin{acknowledgements}
       N.L.I. thanks the National University of La Plata, Argentina - Faculty of Astronomical and Geophysical Sciences of which she is a PhD student. We thank the staff of the GMRT that made these observations possible. GMRT is run by the National Centre for Radio Astrophysics of the Tata Institute of Fundamental Research. This research has made use of the SIMBAD database, operated at CDS, Strasbourg, France.

\end{acknowledgements}
\bibliographystyle{aa}
\bibliography{proplyd}


\end{document}